\documentstyle{mn}

\newcommand{\MDOT}{\rm \dot{M}}
\newcommand{\MSUNYR}{$\rm M_{\odot}\,\rm yr^{-1}$}

\newcommand{\mdot}{\mbox{$\stackrel{.}{\textstyle M}$}}

\title{ A radiation-driven disk wind model for massive young stellar objects.}

\author[J. E. Drew et al.]{Janet E. Drew$^{a}$, Daniel Proga$^{a}$ and 
James M. Stone$^b$\\ 
$\rm ^a$ Imperial College of Science, Technology and Medicine, 
Blackett Laboratory, Prince Consort Road, London SW7 2BZ, UK \\
$\rm ^b$ Department of Astronomy, University of Maryland, College Park 
MD~20742, USA\\
E-mail: j.drew@ic.ac.uk, d.proga@ic.ac.uk and jstone@astro.umd.edu} 

\begin{document}
\maketitle

\begin{abstract}
    A radiation-driven disk wind model is proposed that offers great promise
of explaining the extreme mass loss signatures of massive young stellar
objects (the BN-type objects and more luminous Herbig Be stars).  It is
argued that the dense low-velocity winds associated with young 
late-O/early-B stars would be the consequence of continuing optically-thick
accretion onto them.
The launch of outflow from a Keplerian disk allows wind speeds of $\sim$200 
km s$^{-1}$ that are substantially less than the escape speed from the 
stellar surface.  The star itself is not required to be a rapid rotator.  
Disk irradiation is taken into account in the hydrodynamical calculation
presented, and identified as an important issue both observationally and from 
the dynamical point of view.

\end{abstract}


\section{Introduction} 

     After the first ground-based IR surveys were carried out and as
IR spectroscopy began, it was discovered that some of the
most IR-luminous point sources could be associated with optically-invisible
stars embedded in giant molecular clouds.  The bolometric luminosities of
these objects was determined to be consistent with their being young OB stars
(Wynn-Williams 1982).  Commonly referred to as `BN-type' objects after the 
Becklin-Neugebauer source in OMC-I, these stars were subjected to intensive 
study during the late 1970s and through the 1980s.  

     A key unsolved problem presented by these objects is the dynamical 
origin of the bright, and broad H{\sc i} recombination line emission seen in 
their spectra.  Simon et al. (1983) showed for a number of sources that mass 
loss rates in excess of $10^{-7}$ \MSUNYR are required if the IR line 
emission is attributed to a spherically-symmetric wind.  Field stars of 
comparable (late-O, early-B) spectral type only shed mass at a rate of order 
a few times $10^{-9}$ \MSUNYR (Howarth \& Prinja 1989, see also Cassinelli et
al. 1994).   A further distinction to be made between 
young embedded OB stars and their field counterparts concerns their terminal 
wind speeds: in BN-type objects, 100--300 km s$^{-1}$ is measured from
the IR lines, whereas velocities in the region of 1500 km s$^{-1}$ are
more typical of field main sequence stars (e.g. Prinja, Barlow \& Howarth 
1990).  The low outflow velocities typical of BN-type objects are only
emulated in the field by very low surface gravity supergiants.  The 
apparently high mass loss rates and low outflow speeds have
never been satisfactorily explained, although it has proved possible to 
obtain reasonable fits to observed line flux ratios from models of
spherically-symmetric mass loss (e.g. Simon et al. 1983, H\"oflich \& 
Wehrse 1987).

     Prompted in part by the need to explain the low observed outflow 
velocities and in part by a growing acceptance that all young stars pass
through a disk accretion phase (e.g. Yorke 1986; Adams, Lada \& Shu 1987), 
observations of BN-type
and similar objects began to be interpreted in terms of incomplete disks
and fast winds of undefined origin (e.g. Hamann \& Simon 1986, Persson, 
McGregor \& Campbell 1988).  More recently still, attention has focused on 
the concept of photoevaporation (Hollenbach et al. 1994) which exploits a 
combination of heating and ionization by diffuse radiation and, for more 
luminous exciting stars ($M > 15M_{\odot}$), wind ram pressure in order to 
drive matter away from a remnant protostellar disk.  The motivation for this 
work was to identify a means of fuelling compact and UC H{\sc ii} regions for
long enough to explain the Galactic H{\sc ii} region statistics of Wood \& 
Churchwell (1989).  Significant features of the model are that the main mass 
loss occurs at tens of AU away from the young star and that the stellar
wind itself is normal and, in effect, a boundary condition (see also the 
similar study by Yorke \& Welz 1996).  Our aim is to elucidate the flow on 
this boundary.  

     We argue there is a simple, physically appealing model for the winds
from BN-type and related objects that may yield both the observed IR 
line widths and the seemingly high mass loss rates required.  Specifically, 
we demonstrate that massive YSOs can readily drive 
relatively low-velocity, high-density equatorial winds by means of radiation 
pressure (mediated by line opacity) from Keplerian circumstellar disks that 
reach into the stellar surface.  Indeed, most of the mass is lost from inside
$r \sim 2R_*$. A key feature of the model is that drawing mass from an
extended
Keplerian reservoir breaks the usual link between main sequence high surface 
gravity and high emergent stellar wind speed.  This 
builds upon recent modelling of radiation-driven disk winds applied 
successfully, in the first instance, to cataclysmic variable mass loss 
(Proga, Stone \& Drew 1998, hereafter PSD).  For the model to have relevance 
to BN-type objects and the more luminous Herbig~Be stars, it is sufficient 
that accretion via a circumstellar disk continues after young OB stars 
achieve a near main-sequence configuration.

    In Section 2 we set out the premises of the proposed model, and go on
in section 3 to present the results of a numerical hydrodynamical 
calculation.  The observational consequences of the model are discussed 
in section 4.

\section{The concept}

   Suppose that massive young OB stars spend some time, after
they have acquired a more or less main sequence radius and mass, still
accreting through an equatorial disk.  What would the system look like?  
Consider an early B star ($L \sim 10^4$~L$_{\odot}$, $R \sim 5$~R$_{\odot}$) 
continuing to accrete at a rate of 10$^{-6}$ M$_{\odot}$ yr$^{-1}$ -- a rate 
lying in the range associated with HAeBe stars (Hartmann, Kenyon \& Calvet 
1993), that also shall be seen to be easily high enough to ensure that 
the mass inflow more than replenishes the mass lost through outflow.  The 
accretion luminosity in this case would be entirely negligible compared to 
the stellar luminosity.  Figure~1 presents a rough picture of 
the spectral energy distribution (SED), calculated by assuming that 
blackbody radiation at the appropriate effective temperature is emitted from 
all surfaces.  The accretion component only starts to modify the SED 
noticeably at $\sim2$~$\mu$m.

   Although the self-luminosity of the disk is insignificant, it is 
essential to the model presented here that at the chosen accretion rate,
$\MDOT_{{\rm acc}} \sim 10^{-6}$ M$_{\odot}$ yr$^{-1}$, the disk is likely to 
be optically-thick (see Hartmann et al. 1993).  This means that 
the stellar radiation falling on the disk will be scattered or absorbed and 
then re-emitted, thereby changing the overall geometry of the radiation field.
The apparent SED will also be altered.  First order calculations of this 
effect were carried out by Kenyon \& Hartmann (1987).  In Figure~1, we 
present the result of a calculation like theirs for the case that the 
disk is geometrically thin and completely flat.  We also retain their 
assumptions that the radiation incident on the disk is completely 
thermalised and re-radiated isotropically.  It may be seen in the figure
that the Rayleigh-Jeans tail of the B star's SED is significantly altered by 
the reprocessed component.   

\begin{figure}
\begin{picture}(80,185)
\put(0,0){\includegraphics{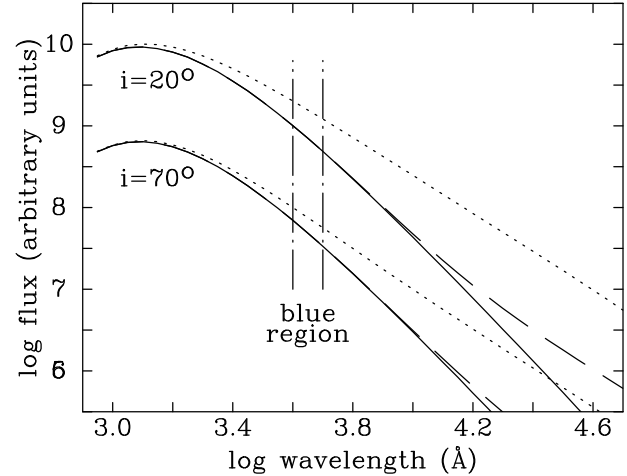}}
\end{picture}
\caption{
Illustrative spectral energy distributions for a 24000~K main sequence
star encircled by an optically-thick accretion disk.  Two sets of curves
have been calculated: one for a nearly pole-on configuration ($i = 20^{o}$),
and the other for a nearly edge-on view ($i = 70^{o}$).  All surfaces are
assumed to radiate isotropically as blackbodies. In each set of 
curves the solid line represents the stellar photosphere only, the
dashed line traces the sum of the stellar photospheric and accretion 
components, while the dotted line also includes the reprocessed component.
The vertical lines mark the blue spectral range normally used to define
spectral type.  A flat disk has been assumed here.  Kenyon \& Hartmann (1987) 
may be consulted for examples of SEDs associated with reprocessing by flared
disks.  
}  
\end{figure}

   The optically-thick disk's reprocessing of the stellar radiation field 
has implications both for observation (a point we shall take up again in 
section 4) and for the outflow dynamics through its effect on the radiation 
field geometry.  Any matter above the disk plane and not so far from the 
stellar surface may see a significant driving flux of radiation from the disk 
as well as directly from the star.  This of course requires that radiation 
pressure is dynamically important. In the immediate vicinity of an early B 
star, radiation pressure mediated by spectral line opacity can be effective 
since $L_*/L_{{\rm Edd}}$ is on the order of 0.01 (for $M \sim$10~M$_{\odot}$ 
and $L \sim 10^4$~L$_{\odot}$), while the opacity presented by plausible 
ensembles of spectral lines can be up to $\sim10^3$ times that 
due to electron scattering alone (Gayley 1995, see also Castor, Abbott \& 
Klein 1975).  Hence there is no physical bar to radiation-driven mass 
loss from a disk around a young early-type star.  Here, in modelling this 
mass loss in the case that the disk is optically-thick, it is assumed that 
all of the direct and reprocessed starlight is effective in driving outflow 
(regardless of its spectral characteristics).

   The perceived difficulty with radiation-driven mass loss from
massive YSOs in the past has been that mass loss direct from the stellar 
photosphere occurs at too high a  velocity and too low a density to make 
sense of the observations. The problem to tackle is how to raise the wind 
density and lower the typical outflow speed.  The solution is rotation.  
Friend \& Abbott (1986) foretold this in their exploration of one-dimensional 
radiation-driven wind solutions admitting a rotating stellar surface as the 
lower boundary .  Significantly,
their solutions indicated markedly decreasing $v_{\infty}$ as the rotation on 
the lower boundary increased.  The case we now describe, in which the wind is 
launched from a reservoir of material in Keplerian orbits 
(i.e. a circumstellar disk), may be thought of as a continuation of this 
trend.

\section{A numerical disk wind model}

\begin{table}
\footnotesize
\begin{center}
\caption{ Full list of model parameters.}
\begin{tabular}{l l  l  } \\ 

\hline
  &  &  \\
Parameter & Value  \\
  &   \\
\hline 
$M_\ast$ &   10~M$_{\odot}$    \\
$r_\ast$ & 5.5~R$_{\odot}$     \\
$L_\ast$ & 8500~L$_{\odot}$ \\
$c_s$    & 14 km s$^{-1}$   \\
$v_{th}$ & 0.3 $c_s$     \\
$k$, $\alpha$       & 0.3, 0.5  \\
$M_{max}$       & 1000     \\
$\rho_{0}$    & $10^{-8}$~g~cm$^{-3}$  \\
$\mdot_{a}$ & $10^{-6}$~M$_{\odot}$~yr$^{-1}$   \\
$r_i $, $r_o$     & 1~$r_\ast$, 10~$r_\ast$    \\
\hline
\end{tabular}
\end{center}
\normalsize
\end{table}

\begin{figure*}
\begin{picture}(170,220)
\put(0,0){\includegraphics{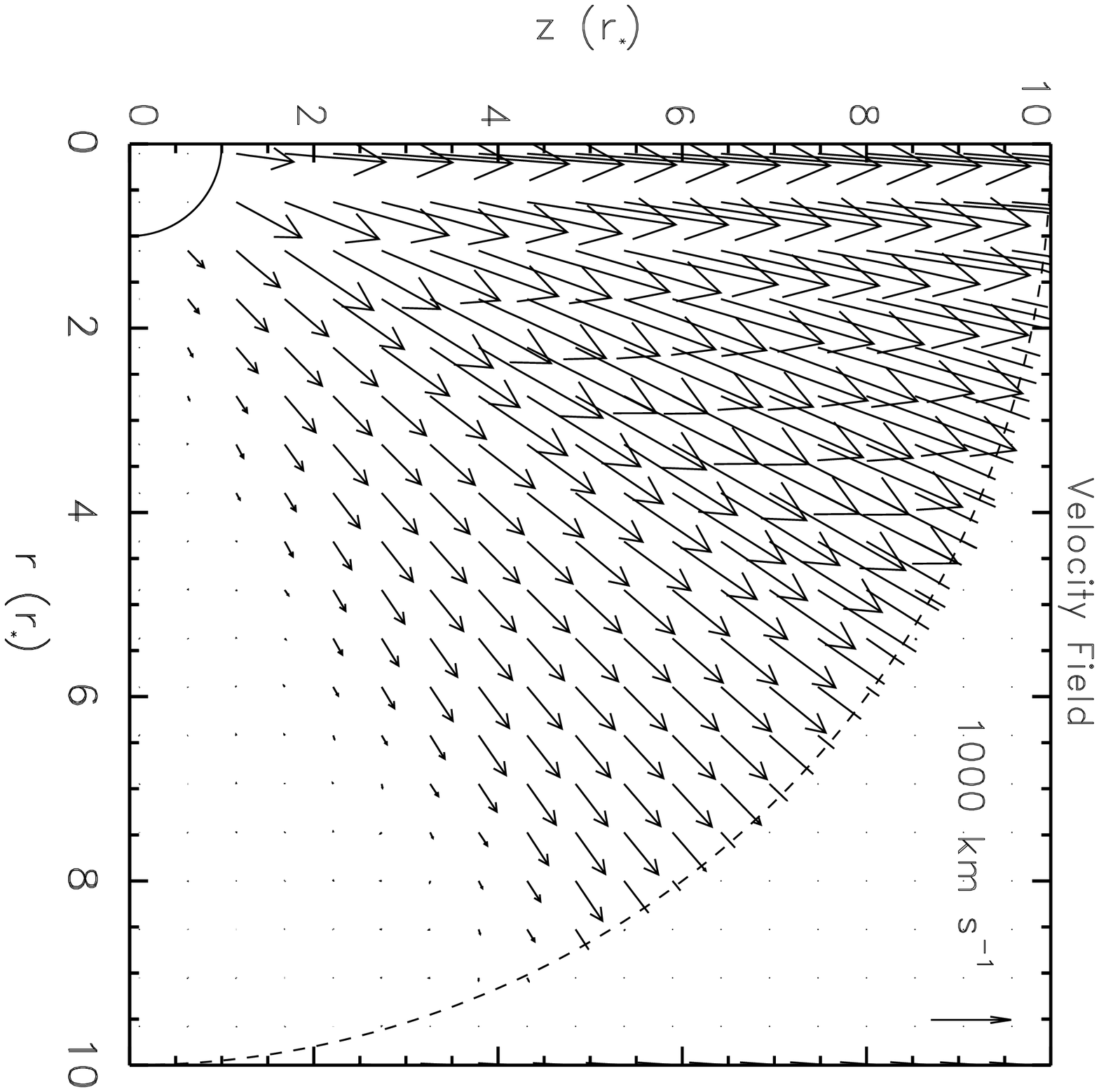}}
\put(0,0){\includegraphics{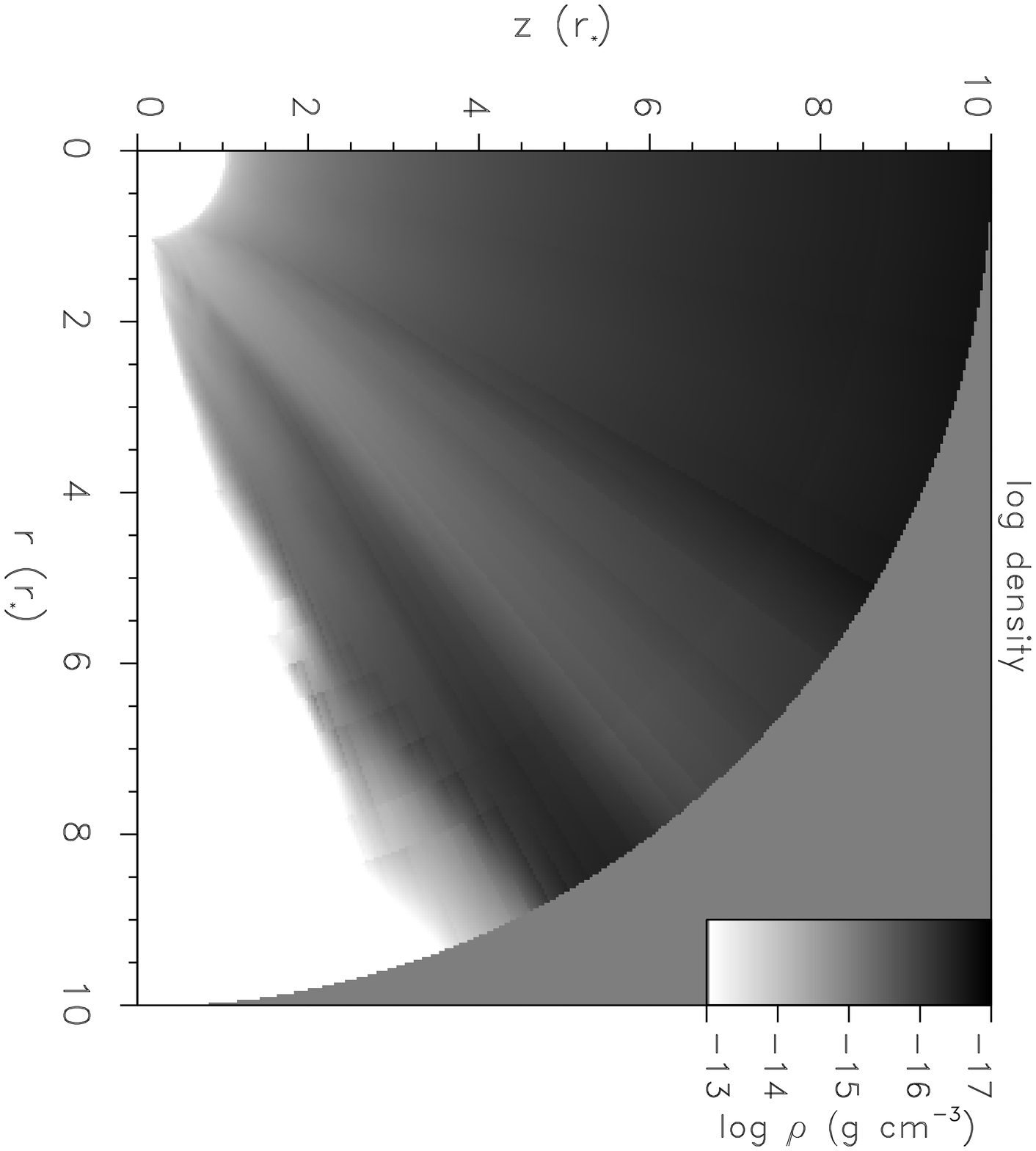}}
\end{picture}
\caption{The left hand side panel is a density map of the early B star disk 
wind model, described in the text. To better illustrate the density changes
in the fast outflow in grey scale, an upper cutoff of 10$^{-13}$~g~cm$^{-3}$ 
has been applied (cf. Figure~3). 
The right hand side panel is a map of the velocity field (the poloidal 
component only). 
In both panels the rotation axis of the star is along the left hand 
vertical frame, while the mid-plane of the circumstellar accretion star is 
along the lower horizontal frame. 
}
\end{figure*}

   Our 2.5-dimensional hydrodynamical numerical method is in most respects
as described by PSD.  However, here, mass is allowed to enter the 
computational domain from the star as well as from the disk.  In order to 
resolve the subsonic portion of the flow from the star it was necessary to 
increase the radial grid resolution by using a zone size ratio of 1.08 
instead of 1.05 (cf. PSD).  The entire domain is sampled by 100$\times$100 
points.  We consider the case of a non-rotating star of mass 10~M$_{\odot}$, 
radius 5.5~R$_{\odot}$ and luminosity $8500$~L$_{\odot}$, accreting at a rate 
of 10$^{-6}$ M$_{\odot}$ yr$^{-1}$.  Note that this implies 
$L_\ast/L_{{\rm acc}} = 300$ ($x = 300$ in the notation of PSD) and hence
that reprocessed starlight overwhelmingly dominates the circumstellar disk's
light output.  The disk is assumed to be flat.  A full list of model 
parameters is given in Table 1.

   Our calculation of the radiation force (per unit mass) due to spectral 
lines at every location in the flow is of the form
\begin{equation}
    F^{rad,l} = \int \left(\frac{\sigma_e d{\cal F}}{c}\right) M(t) .
\end{equation}
The term in brackets is the electron scattering radiation force and $M$, 
the force multiplier, is the increase in the scattering cross section due to 
line opacity.  The integration is over all visible radiating surfaces.  
Presently, the force multiplier used is of the simple form $M(t) = 
kt^{-\alpha}$, where $k$ and $\alpha$ are constants and $t$ is proportional 
to the local density divided by the relevant velocity gradient.  The 
complicated interplay between the geometry and flow kinematics demands 
careful evaluation of the radiation force integral (see PSD).
We adopt $k = 0.3$, $\alpha = 0.5$ and impose an upper limit 
$M_{max} = 10^3$ on the force multiplier.  The values picked
for $k$ and $\alpha$ have been guided by the literature on line-driving of 
early B star winds (e.g. de Ara\'ujo, de Freitas Pacheco \& Petrini 1994; 
Lamers, Snow \& Lindholm 1995).    In the early B spectral 
domain ($T_{eff} \sim 20000$~K), it is known that the radiation force exhibits
a `bistability' in the sense that low Lyman continuum opacity ($\tau$ less 
than a few) yields a more highly ionized medium in which a smaller number of 
optically-thick lines dominate the force, while higher Lyc opacity produces 
lower overall ionization and driving by a larger number of optically-thin 
transitions (Lamers \& Pauldrach 1991).  In the interests of simplicity, the 
present parameterisation of the force does not make this distinction but 
instead uses values of $k$ and $\alpha$ that are intermediate between these 
two regimes.  

    Everywhere, we set the sound speed, $c_s$, to be 14 $\rm km~s^{-1}$, a 
value that corresponds to T $\sim$ 15000~K, a temperature that is plausible 
for ionized gas in close proximity to an early B star (see Drew 1989; the 
precise value of $T_{eff}$ in our model is 23660~K).  The sound speed, in 
combination with the local gravity, determines the pressure scale height in 
the atmospheres of both the star and 
disk.  The boundary density, $\rho_0$, along the stellar and disk surfaces is 
constant in time and set to $\rho_0~=~10^{-8}~$
g~cm$^{-3}$.

    Figure 2 shows the converged model density distribution and 
velocity field.  A steady state is achieved, in which the calculation follows 
(i) a fast polar wind from the star, (ii) a transitional zone in which the 
stellar wind experiences some streamline compression due to the presence of 
the disk flow, and is exposed to non-radial line-driving due to the
reprocessed disk radiation, (iii) a very much denser, slower equatorial 
outflow from the disk.  In the present case, where it has been 
assumed that the disk is optically thick and reprocesses all incident stellar 
light into dynamically useful, isotropically re-emitted, light, the surface of
the denser, slow equatorial flow lies at $\theta \sim 60^o$ (where $\theta$ 
is the colatitude angle measured with respect to the polar direction).

    The polar wind densities and outflow speeds obtaining within the model
shown in figure 2 are very much in line with modified CAK theory.  Without 
the disk, the total mass loss rate in the spherically-symmetric wind would 
be $\sim10^{-8}$ M$_{\odot}$ yr$^{-1}$ and the terminal velocity, $v_*$ would 
be $\sim$2000 km s$^{-1}$.  The mass loss rate in the model shown is 
dominated by the disk component and is $\sim$3 times higher.  However, the 
density contrast between the polar wind and the peak of the mass flux at an 
angle of $\sim$70$^o$ is over two orders of magnitude (see Figure 3)!  
Radial velocities in the stream leaving the disk range from 400 km~s$^{-1}$ 
down to zero at the de facto boundary between the disk atmosphere and 
outflow (Figure 3 also).  The comparison that the measured H{\sc i}-line 
FWHM in BN-type objects and Herbig~Be stars are typically a couple of hundred 
km s$^{-1}$ or so is thus very encouraging.  Currently, the disk atmosphere 
scale height towards larger disk radii is undoubtedly exaggerated by the 
isothermal approximation made within the model.  Tests suggest that when this 
is corrected there is unlikely to be much change in either the mass loss rate 
or expansion velocities.  However the disk component of the wind would very 
likely be more markedly equatorial than here.

\begin{figure}
\begin{picture}(100,130)
\put(0,0){\includegraphics{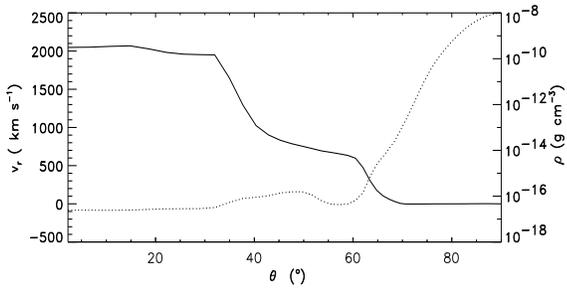}}
\end{picture}
\caption{Quantities at the outer boundary in the model from Figure 2.
The ordinate on the left hand side is marked by the solid line, while
the ordinate on the right hand side is marked by the dotted line. 
}  
\end{figure}
\section{ Implications of the model }

    The promise of this model is that it provides, in a conceptually simple
way, an appropriately high density, slowly expanding outflow.  An order
of magnitude estimate of the enhancement in the H{\sc i} line emission 
measure ($\int n_e^2 dV$) with respect to a normal early-B main sequence 
stellar wind can be derived.  To do this, it is assumed that the normal
spherical wind and the disk wind are both uniform within the volumes
occupied and that the radial acceleration of both is approximately the
same.  The scaling then is as:
\begin{equation}
   (\frac {\MDOT_d}{\MDOT_*})^2(\frac{v_*}{v_d})^2(\frac{1}{\cos \theta_d})
\end{equation}
where the subscripts distinguish the normal spherically-symmetric stellar
wind from the dense, slow disk wind.  Crudely speaking, the angle $\theta_d$ 
is the angle between the `surface' of the disk outflow and the stellar pole.
Inserting the quantities already mentioned above and adopting
a typical disk outflow speed of $v_d \sim 200$ km~s$^{-1}$, one obtains
an emission measure enhancement in excess of 1000.  

    Is this enough?  The same expression can be used to scale with 
respect to published H{\sc i} line models of BN-type objects.
For the moment, assume the ratio of velocities is of order unity.  The fits 
to radio and H{\sc i} line data performed by Simon et al. (1983) and by 
H\"oflich \& Wehrse (1987) assumed spherical symmetry and constant velocity 
outflow, and derived mass loss rates within a factor of a few of 
$5\times10^{-7}$ M$_{\odot}$~yr$^{-1}$.  Comparison between this mass loss 
rate and the disk wind mass loss rate obtained here would suggest that up to a 
factor of 100 in emission measure is still missing.  At this point, however, 
it has to be acknowledged that there are significant complexities in
excitation and geometry that the naive scaling fails to quantify.  Ahead 
of synthesising spectral line profiles from the model, it is unclear just what 
`representative' disk wind velocity should be adopted in the scaling.  
H{\sc i} line emission will undoubtedly be optically-thick (as they are 
observed to be), which begs the question as to how radiation transfer effects 
will define the line-forming region.   Nevertheless, it is reasonable
to anticipate that the disk-like geometry, with its steep latitudinal density 
gradients, would favour emission from material at a higher mean square 
density than would a $\sim 1/r^2$ spherically-symmetric density profile.  

    There is every prospect that outflow rather than rotation will 
dominate the shape of optically-thick spectral lines.  If so, this
will fit in with the typically single-peaked appearance of H{\sc i} line 
profiles in the spectra of massive YSOs (e.g. Bunn, Hoare \& Drew 1995).  
Since angular momentum conservation will apply, the rotational component of 
motion is only comparable with the outflow velocities achieved within the 
disk atmosphere and close to the stellar surface. Only optically-thin lines 
originating close to the disk plane, perhaps within neutral or at-best 
partially ionized gas, may produce double-peaked line profiles.  For the
time being, it is already encouraging that the relatively equatorial 
configuration we have identified has moved in the direction of the findings of 
recent high spatial resolution studies of S140~IRS1 and S106IR (Hoare et al. 
1994; Hoare, Glindemann \& Richichi 1996; Hoare \& Muxlow 1996).  After
the inclusion of an energy balance constraint, in place of the present
isothermal approximation, the model outflow is likely to be even more
equatorial.

    Is the scenario proposed here the only one with a chance of providing
a workable equatorial stellar wind model?  Until recently there was a
prospect that the wind-compressed disk concept of Bjorkman \& Cassinelli 
(1993) would apply to massive YSOs -- just as it might apply to B[e] and
classical Be stars.  Within the last year or so Owocki, Cranmer \& Gayley
(1996) have
shown that the detailed velocity-field dependence of the radiation force
inhibits the equatorial focusing of the flow envisaged by the model.  Even 
before the wind compressed disk model appeared, it had been proposed that 
`bistability' (Lamers \& Pauldrach 1991) would, in the case of rapidly 
rotating stars, induce a dense, slow equatorial wind in combination with a 
fast low-density polar wind.  Our model achieves a density contrast of
up to two orders of magnitude, even in the case of a negligibly-rotating star 
and a uniform force multiplier ($k=0.3$ and $\alpha=0.5$ everywhere).  
Here the slow, dense equatorial flow is a consequence of its being drawn from 
gas in orbit around the star rather than from the star's surface.  
`Bistability', if it were also accounted for in the disk wind model, may
enhance the density contrast between the polar and equatorial flows a
little further -- for $L_*/L_{{\rm Edd}} \sim 0.01$, the scaling formula 
given by Owocki (1997) limits this factor to about 3.

    A significant element in our model is the irradiation of the 
optically-thick circumstellar disk by the star (Figure~1). 
Our simple model of the SED predicts that direct starlight and the reprocessed
component should be comparable in the blue part of the spectrum, while the
latter dominates longward of 1$\mu$m.  The hotter direct stellar component 
only takes over towards its Planck maximum in the ultraviolet.
Since a relatively large fraction of the optical light may be attributable to 
the disk, heavy element absorption due to the stellar photosphere could prove 
to be hard to detect.  The disk itself will at best contribute 
rotationally-smeared, weaker line absorption or may indeed produce only line 
emission (given that the run of temperature vertically within the disk 
induced by the strong external irradiation will be inverted; see Hubeny 
1990).  Detailed theoretical models have yet to be constructed and so we 
should consider what can be learned from the observations that, only now, are 
beginning to be gathered, 
Recent repeat observations (Oudmaijer et al. in preparation) of the
young B1.5 star, MWC~297,  have revealed changes in the $B$-band heavy 
element absorption lines with respect to previous observations (Drew et al.  
1997) that show they are not simply photospheric.  In a study of young stars 
in M17, Hanson, Howarth \& Conti (1997) failed altogether to detect any 
heavy-element photospheric features between 4100\AA\ and 4800\AA\ in their 
putative early B stars.  The first evidence therefore supports the idea
that the optical spectra of high mass YSOs are veiled.

    Finally, we turn to the larger context of this work.  There can be a 
fundamental physical distinction between massive and lower mass YSO winds: 
radiation pressure may take the role in driving mass loss from the 
former that only MHD effects can assume in the latter.  It is perhaps 
significant that highly-collimated jets are a phenomenon more obviously 
associated with lower luminosity YSOs, than with BN-type objects.  Certainly, 
molecular bipolar flows are associated with S106IR and several other high 
luminosity YSOs, but there are no compelling examples of the ionized tight 
jets to compare with e.g. the HH~34 system (see Bally 1997).  It is generally
believed that such extreme collimation requires effective magnetic hoop 
stresses (see the review by K\"onigl \& Ruden 1993, but cf. Mellema \&
Frank 1998).  The angular-momentum 
conserving equatorial wind that we have proposed here does not favour
the growth of the necessary toroidal magnetic field.  Nevertheless, it would
be premature to suggest that magnetic fields have no role to play in
young massive stars.  That remains to be seen.  Our contention here is that 
radiation pressure must have a role and that now it is possible to 
describe its action in an axially-symmetric context that suggests a promising 
model for these objects.

{\bf Acknowledgments:} This research has been supported by a research
grant from PPARC, and by NASA through HST grant GO-6494.  Computations
were performed at the Pittsburgh Supercomputing Center.


\begin{thebibliography}{}
\bibitem[]{}
  Adams F.C., Lada C.J., Shu F.H., 1987, ApJ, 312, 788
\bibitem[]{}
  Bally J., 1997, in {\sl Accretion phenomena and related outflows}, eds.
  D.T. Wickramasinghe, G.V. Bicknell \& L. Ferrario, ASP Conf.Sers. 121, 3
\bibitem[]{}
  Bjorkman J.E., Cassinelli J.P., 1993, ApJ, 409, 429
\bibitem[]{}
  Bunn, J.C., Hoare, M.G., Drew, J.E., 1995, MNRAS, 272, 346
\bibitem[]{}
  Cassinelli J.P., Cohen D.H., Macfarlane J.J., Sanders W.T., Welsh B.Y.,
  1994, ApJ, 421, 705
\bibitem[\protect\citename{CAK}1975]{CAK} 
  Castor J.I., Abbott D.C., \& Klein R.I. 1975, ApJ, 195, 157 (CAK)
\bibitem[]{}
  de Ara\'ujo F.X., de Freitas Pacheco J.A., Petrini D., 1994, MNRAS, 267,
  501
\bibitem[]{}
  Drew J.E., 1989, ApJS, 71, 267
\bibitem[]{}
  Drew J.E., Busfield G., Hoare M.G., Murdoch K.A., Nixon C.A., Oudmaijer 
  R.D., 1997, MNRAS, 286, 538 
\bibitem[\protect\citename{Friend \& Abbott}1986]{fa}
  Friend D.B., Abbott D.C. 1986, ApJ, 311, 701
\bibitem[\protect\citename{Gayley}1995]{ga}
  Gayley K. G. 1995, ApJ, 454, 410
\bibitem[\protect\citename{Hamann \& Simon}1986]{hs}
  Hamann F., Simon M., 1986, ApJ, 311, 909
\bibitem[]{}
  Hanson M.M., Howarth I.D., Conti P.S., 1997, ApJ, in press
\bibitem[]{}
  Hartmann L., Kenyon S.J., Calvet N., 1993, ApJ, 407, 219
\bibitem[]{}
  Hoare M.G., Drew J.E., Muxlow T.B., Davis R.J., 1994, ApJ, 421, L51
\bibitem[]{}
  Hoare M.G., Glindemann A., Richichi A., 1996, in {\sl The role of dust in
  the formation of stars}, ESO Conf.Proc., Springer Verlag, p35
\bibitem[]{}
  Hoare M.G., Muxlow T. B., 1996, in {\sl Radio emission from the stars and
  Sun}, ASP Conf.Sers. Vol.93, p47
\bibitem[]{}
  H\"oflich P., Wehrse R., 1987, A\&A, 185, 107
\bibitem[]{}
  Hollenbach D., Johnstone D., Lizano S., Shu F., 1994, ApJ, 428, 654
\bibitem[]{}
  Howarth I.D., Prinja R.K. 1989 ApJS, 69, 527 
\bibitem[]{}
  Hubeny I., 1990, ApJ, 351, 632
\bibitem[]{}
  Kenyon S.J., Hartmann L., 1987, ApJ, 323, 714
\bibitem[]{}
  K\"onigl A., Ruden S.P., 1993, in {\sl Protostars and Planets III},
  eds. E. H. Levy \& J. I. Lunine, Tucson: U. Arizona Press, 641 
\bibitem[]{}
  Lamers H.J.G.L.M., Pauldrach A.W.A., 1991, A\&A, 244, L5
\bibitem[]{}
  Lamers H.J.G.L.M., Snow T.P., Lindholm D.M. 1995, ApJ, 455, 269
\bibitem[]{}
  Mellema G., Frank A., 1998, MNRAS, 292, 795
\bibitem[]{}
  Owocki S.P., Gayley K.G., Cranmer S.R., 1998, in {Properties of hot
luminous stars}, ed. I. D. Howarth, ASP Conf.Sers., in press
\bibitem[]{}
  Owocki S.P., Cranmer, S.R., Gayley K.G., 1996, ApJ, 472, L115 
\bibitem[]{}
  Persson S.E., McGregor P.J., Campbell B., 1988, ApJ, 326, 339
\bibitem[]{}
  Prinja R.K., Barlow M.J., Howarth I.D., 1990, ApJ, 361, 607
\bibitem[]{}
  Proga D., Stone J.M., Drew J.E. 1998, MNRAS, in press (PSD)
\bibitem[\protect\citename{Simon et al. }1983]{si} 
  Simon M., Felli M., Cassar L., Fischer J., Massi M., 1983, ApJ, 266, 623
\bibitem[]{}
  Yorke H. W., 1986, ARAA, 24, 49
\bibitem[]{}
  Yorke H. W., Welz A., 1996, A\&A, 315, 555
\bibitem[]{}
  Wood D.O.S.W., Churchwell E., 1989, ApJS, 69, 831
\bibitem[]{}
  Wynn-Williams C.G., 1982, ARAA, 20, 587
\end{thebibliography}
\end{document}